\documentclass[iop,apj,12pt]{emulateapj}
\usepackage{times}
\usepackage[normalem]{ulem}
\usepackage{epsfig}
\usepackage{natbib}
\usepackage{amsfonts}
\usepackage{amsmath}
\usepackage{multirow}
\usepackage{enumerate}
\usepackage[usenames]{color}
\usepackage[plainpages=false, colorlinks=true, anchorcolor=blue, linkcolor=blue, citecolor=blue, bookmarks=false]{hyperref}
\newcommand{\be}{\begin{eqnarray}}
\newcommand{\ee}{\end{eqnarray}}
\newcommand{\lp}{\left(}
\newcommand{\rp}{\right)}

\newcommand{\dens}{\rho_5}
\newcommand{\temp}{T_8}

\newcommand{\spin}{\Omega_{0.1}}

\begin{document}


\slugcomment{Accepted for publication in The Astrophysical Journal}

\title{Turbulent Mixing on Helium-Accreting White Dwarfs}

\author{Anthony L. Piro}

\affil{Carnegie Observatories, 813 Santa Barbara Street, Pasadena, CA 91101, USA; piro@obs.carnegiescience.edu}


\begin{abstract}
An attractive scenario for producing Type Ia supernovae (SNe Ia) is a double detonation, where detonation of an accreted helium layer triggers ignition of a C/O core. Whether or not such a mechanism can explain some or most SNe Ia depends on the properties of the helium burning, which in turn is set by the composition of the surface material. Using a combination of semi-analytic and simple numerical models, I explore when turbulent mixing due to hydrodynamic instabilities during the accretion process can mix C/O core material up into the accreted helium. Mixing is strongest at high accretion rates, large white dwarf (WD) masses, and slow spin rates. The mixing would result in subsequent helium burning that better matches the observed properties of SNe Ia. In some cases, there is considerable mixing that can lead to more than 50\% C/O in the accreted layer at the time of ignition. These results will hopefully motivate future theoretical studies of such strongly mixed conditions. Mixing also has implications for other types of WD surface explosions, including the so-called .Ia supernovae, the calcium-rich transients (if they arise from accreting WDs), and metal-enriched classical novae.
\end{abstract}

\keywords{accretion, accretion disks ---
	hydrodynamics ---
	white dwarfs}


\section{Introduction}

The use of Type Ia supernovae (SNe Ia) as cosmic distance indicators \citep[e.g.,][]{Riess98,Perlmutter99} has put increased emphasis on the theoretical uncertainties that still remain about these events. It is generally accepted that SNe Ia result from the unstable thermonuclear ignition of a degenerate C/O white dwarf (WD), but the specific progenitor systems have not yet been identified. The two main candidates can roughly be divided into (1) stable accretion from a non-degenerate binary companion until the Chandrasekhar limit is reached \citep[single degenerates,][]{Whelan73}, (2) the merger of two C/O WDs \citep[double degenerates,][]{Iben84, Webbink84}. An important problem is therefore to understand which scenarios explain the SNe Ia that we observe and whether any one channel is dominant.

How these systems ignite is also  an important outstanding question. One ignition scenario that could occur in both single and double degenerate systems is the ``double detonation.'' This happens when accretion and detonation of a helium shell on a C/O WD leads to a prompt detonation of the core \citep{Woosley94,Livne95}. Double detonations have a number of attractive features which has led to increased interest in this scenario in recent years. Observationally, they are consistent with limits on shock cooling of the exploding WD \citep{Piro10,Bloom12}, the non-detection of a companion in pre-explosion imaging of nearby SNe Ia \citep{Li11}, the lack of radio emission \citep{Hancock11,Horesh12}, the lack of hydrogen emission in nebular spectra \citep{Leonard07,Shappee13}, a lack of a signature of ejecta interaction with a companion \citep{Kasen10,Hayden10},  the missing companions in SNe Ia remnants \citep{Schaefer12}, and the sub-Chandrasekhar ejecta masses inferred in many cases \citep{Scalzo14}. On the theoretical side, the sub-Chandrasekhar masses of the C/O cores when they ignite can roughly explain the size of the width-luminosity relation \citep{Phillips93} observed for typical SNe Ia \citep{Sim10} and population synthesis calculations argue that double detonations could explain the majority of normal SNe Ia \citep{Ruiter11,Ruiter14}. Substantial theoretical work has also focused on the details of the helium and/or C/O burning \citep{Fink07,Fink10,Moll13}.

The main issues with the double detonation scenario have been whether the helium ignition is robust \citep{Holocomb13} and whether the colors and spectra due to the ashes left over from the surface helium burning are strongly inconsistent with normal SNe Ia \citep{Woosley11,Sim12}. However, it was recently argued that both of these difficulties may be overcome by if there is additional C/O material in the surface helium layer prior to when the detonation begins \citep[][the latter of which also explored a more detailed treatment of the nuclear network]{Kromer10,Shen14b}. The main argument is that the additional burning decreases the hotspot size and mass needed for triggering a detonation, which in turn leads to less iron-peak elements in the helium layer that adversely affect the spectra. In addition, the silicon and calcium left over from the surface burning may lead to high-velocity spectral features as observed in most SNe Ia \citep{Mazzali05,Tanaka08,Blondin12,Childress14,Maguire14}. This C/O can get created in the surface layer by the convective helium burning in the time prior to initiation of the detonation \citep{Shen09} or mixed in from the C/O core by convective mixing associated with this same stage. Here I consider in more detail whether C/O material can also be mixed into the helium layer via turbulent mixing during the accretion process.

Additional motivations for this work are the scenarios where the C/O core is not successfully ignited by the surface helium detonation \citep{Shen14}. In such cases, the low mass of the helium would lead to a faint and rapidly evolving transient \citep{Bildsten07,Shen10,Waldman11}, which may be related to various recently observed rapid transients \citep{Perets10,Kasliwal10,Kasliwal12,Foley13,Inserra15}. Whether or not a helium shell-burning scenario can explain a given observation will depend on the composition of the surface layers, which again can be affected by turbulent mixing.

The impact of spin on helium-accreting WDs has been the subject of previous theoretical work \citep{Yoon04b,Yoon04c}. These studies were mainly focused on the shear instabilities for providing the angular momentum transport, viscous heating, and material mixing. Since I show below that shear instability is likely subdominant to other processes, it is worth revisiting these models. Furthermore, new questions about the double detonation scenario and new classes of optical transients (both described above) mean that it is useful to address mixing in the context of these new potential applications.

In Section \ref{sec:equations}, I summarize the picture of turbulent mixing and angular momentum transport that I utilize. In Section \ref{sec:transport}, I discuss hydrodynamic mixing mechanisms and analytic scalings which demonstrate how it depends on accretion rate and spin. In Section \ref{sec:numerical}, I compare with numerical models to demonstrate these main features. I conclude in Section \ref{sec:conclusions} with a summary of my results and a discussion of future work.


\begin{figure*}
\centering
\includegraphics[width=0.8\textwidth]{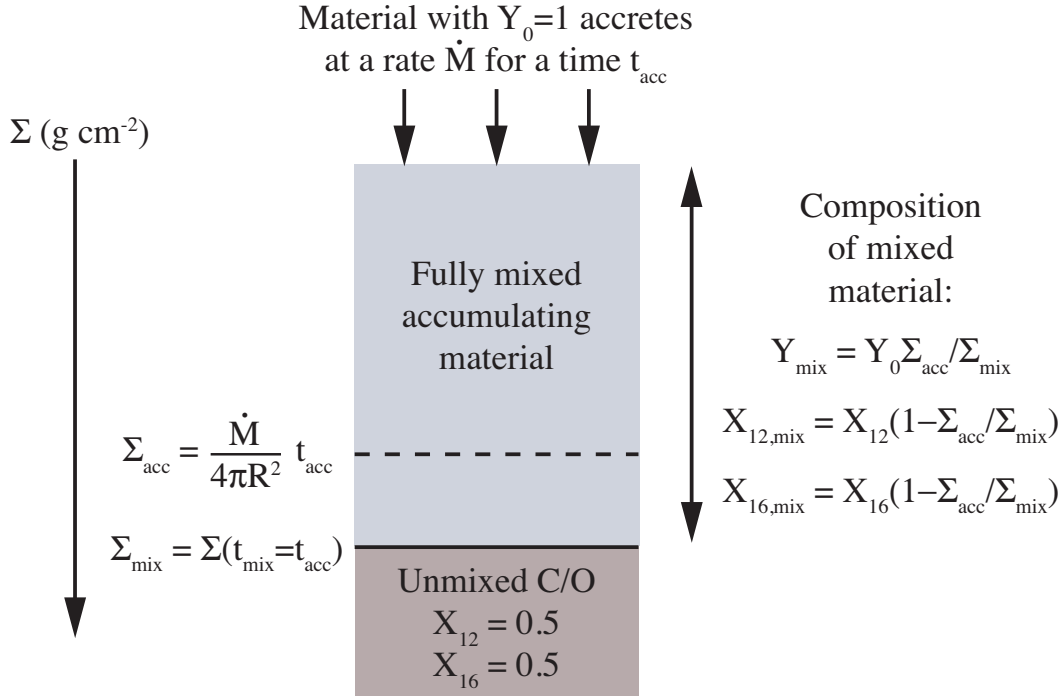}
\caption{Diagram summarizing the main features of mixing considered here. Material is accreted for a timescale $t_{\rm acc}$ at a rate $\dot{M}$, which corresponds to a column depth $\Sigma_{\rm acc}=\dot{M}t_{\rm acc}/4\pi R^2$. As long as $t_{\rm mix}<t_{\rm acc}$ at $\Sigma_{\rm acc}$, material is mixed down to the depth where $t_{\rm mix}=t_{\rm acc}$. This corresponds to a column of $\Sigma_{\rm mix}>\Sigma_{\rm acc}$. The surface layer becomes completely mixed for all depths shallower than this (much like throughout the convective region of a star), and thus it has a helium mass fraction of $Y_{\rm mix}=Y_0\Sigma_{\rm acc}/\Sigma_{\rm mix}$, where $Y_0$ is the helium mass fraction in the freshly accreted material. Carbon and oxygen are also mixed up into the surface layer and have corresponding mass fractions as specified in the diagram.}
\label{fig:diagram}
\end{figure*}

\section{Transport and Mixing Framework}
\label{sec:equations}

Before moving into the main calculations, it is helpful to first explain the general strategy utilized by this work. Both the treatment of the angular momentum transport and the turbulent mixing are somewhat different than what is typically used, so it is helpful to explain why this approach is useful. Similar methods  were used in \citet{Piro07}, and thus readers interested in additional details should consult this work.

\subsection{Angular Momentum Transport}

Material accreting onto a WD at a rate $\dot{M}$ reaches its surface with a nearly Keplerian spin frequency
\be
	\Omega_{\rm K} = (GM/R^3)^{1/2} = 1.0M_1^{1/2}R_{8.7}^{-3/2}\,{\rm s^{-1}},
	\label{eq:omegak}
\ee
where $M_1=M/1\,M_\odot$ and $R_{8.7}=R/5\times10^8\,{\rm cm}$ are the typical WD mass and radius I will be considering. Most of the energy of this incoming material is dissipated in a thin boundary layer and does not reach far into the star \citep{Piro04}. Nevertheless, this material adds angular momentum at a rate of $(GMR)^{1/2}\dot{M}$, and so a torque of this magnitude must be present to transport the angular momentum through the WD and increase its spin.

Since these shallow surface layers of the WD have a typical pressure scale height $H\ll R$, where $H=P/\rho g$ and $g=GM/R^2$ is the surface gravitational acceleration, I assume a plane-parallel geometry and set $R$ as the radius and use $z$ as the vertical height coordinate above $R$. This also allows me to introduce a useful variable in the column depth
\be
	d\Sigma = -\rho dz,
\ee
to measure the depth in the atmosphere. In the plane-parallel limit, the pressure given by hydrostatic equilibrium is simply $P=\Sigma g$. In this geometry, the transport of angular momentum is described by a one-dimensional diffusion equation \citep{Fujimoto93}
\be
	\frac{d}{dt}\lp R^2\Omega\rp = \frac{1}{R^2\rho}\frac{\partial}{\partial z} \lp \rho\nu R^4 \frac{\partial \Omega}{\partial z} \rp,
	\label{eq:angularmomentum}
\ee
where $\Omega$ is the spin of the WD and $\nu$ is the viscosity for transporting angular momentum. In principle $\nu$ could be something like a molecular diffusivity, but this is never relevant for the cases considered here and I instead focus on $\nu$ as a turbulent diffusivity.

Equation (\ref{eq:angularmomentum}) can be significantly simplified by assuming steady-state transport of angular momentum \citep[also see][]{Piro08}. This is reasonable as long as the angular momentum diffusion timescale at a given depth
\be
	t_{\rm visc} = H^2/\nu,
\ee
is less than the timescale to accrete down to that same depth
\be
	t_{\rm acc} = 4\pi R^2\Sigma/\dot{M},
\ee
which is easily satisfied in the WD surface layers. The total time derivative then simplifies to $d/dt \approx V_{\rm adv} \partial/\partial z$, where $V_{\rm adv}=-\dot{M}/4\pi R^2\rho$ is the advecting velocity of the fluid in an Eulerian frame. Making this substitution, integrating with respect to $z$, and taking the limit $\Omega\ll \Omega_{\rm K}$, the final result is
\be
	\sigma = \frac{\dot{M}\Omega_{\rm K}}{4\pi R\rho\nu},
	\label{eq:angularmomentum2}
\ee
where $\sigma=d\Omega/d\ln r$ is the shear rate in the WD. Note that this can also be rewritten in the dimensionless form
\be
	\frac{\sigma}{\Omega} = \lp \frac{t_{\rm visc}}{t_{\rm acc}} \rp \lp \frac{R}{H} \rp \lp \frac{\Omega_{\rm K}}{\Omega} \rp.
\ee
This shows that a non-zero shear must exist throughout the star to allow the star to spin up and that it is {\it larger} when the viscosity is small (so that $t_{\rm visc}$ is big), the accretion rate is large (so that $t_{\rm acc}$ is small), the thickness $H$ of the layer is smaller, or when the spin $\Omega$ is smaller.

\subsection{Turbulent Mixing Framework}
\label{sec:mixing framework}

Traditionally mixing is modeled as a diffusive process where for some element $i$ with mass fraction $X_i$,
\be
	\frac{d X_i}{d t} =  \frac{1}{r^2\rho} \frac{\partial}{\partial r} \lp r^2\rho D\frac{\partial X_i}{\partial r} \rp,
\ee
where $\rho$ is the density and $D$ is the diffusion coefficient for mixing, which will vary with depth in the star. While this is well-suited for implementation in one-dimensional stellar evolution codes \citep[e.g.,][]{Heger00}, an alternative way to think about the mixing is like convection. Just as a region in the star is either convectively mixed or not, a region in the star can be turbulently mixed or not.

The general picture I employ is summarized in \mbox{Figure \ref{fig:diagram}} and described next. After accreting at a rate $\dot{M}$ for a timescale $t_{\rm acc}$, the column depth at the base of the accretion is
\be
	\Sigma_{\rm acc} = \frac{\dot{M}t_{\rm acc}}{4\pi R^2},
\ee
as shown by the horizontal dashed line in Figure \ref{fig:diagram}. The turbulent mixing time at this depth is estimated as
\be
	t_{\rm mix}(\Sigma_{\rm acc}) = H^2/D,
	\label{eq:t_mix}
\ee
where the right-hand side is evaluated at $\Sigma_{\rm acc}$. If  $t_{\rm mix}(\Sigma_{\rm acc})>t_{\rm acc}$, then the mixing takes too long to act in comparison to the accretion and there is minimal mixing. If instead $t_{\rm mix}(\Sigma_{\rm acc})<t_{\rm acc}$, then the mixing can carry material to depths larger than the accretion depth.

Evaluating $t_{\rm mix}$ at every depth in the star, one can then find the depth where $t_{\rm mix}=t_{\rm acc}$, which is denoted $\Sigma_{\rm mix}$. This is the depth down to where material is mixed, as shown in light gray in Figure \ref{fig:diagram}. I assume everything is uniformly mixed down to this depth, which is roughly correct because the mixing timescale $t_{\rm mix}$ is generally smaller at shallower depths. Thus if something mixes just a little at the base of the mixed region, the mixing is fast enough to mix it everywhere above this depth. In a sense, this is analogous to the widely used assumption of complete mixing in the convective regions of stars. Since material is mixed down to $\Sigma_{\rm mix}$, the mass fraction of helium in the mixed layer is decreased to be
\be
	Y_{\rm mix} = Y_0 \Sigma_{\rm acc}/\Sigma_{\rm mix},
\ee
where $Y_0$ is the initial helium fraction of the material when it is first accreted (this variable is introduced just for completeness, I always use $Y_0=1$). Similarly, the C/O, which have mass fractions of $X_{12}$ and $X_{16}$ in the C/O core, will be mixed up and have mass fractions
\be
	X_{12,\rm mix} = X_{12}(1-\Sigma_{\rm acc}/\Sigma_{\rm mix}),
\ee
and
\be
	X_{16,\rm mix} = X_{16}(1-\Sigma_{\rm acc}/\Sigma_{\rm mix}).
\ee
in the mixed layer.

\section{Turbulent Transport Mechanisms}
\label{sec:transport}

To investigate the angular momentum and material transport discussed above requires setting $\nu$ and $D$, respectively. For the present work I focus on transport mediated by hydrodynamic instabilities triggered by accretion at the WD surface. I next summarize the Kelvin-Helmholtz instability and the baroclinic instability since these traditionally are most effective at radial transport (rather than latitudinal transport like from Eddington-Sweet circulation). I hold off on considering magnetohydrodynamic instabilities for future work. This would likely be something like the Tayler-Spruit dynamo \citep{Spruit02}, although this may overestimate the turbulent viscosity in some cases \citep[e.g.,][]{Cantiello14}.

\subsection{Kelvin-Helmholtz Instability}

   The Kelvin-Helmholtz instability (also referred to as the dynamical shear instability) is governed by the Richardson number
\be
	Ri \equiv \frac{N^2}{\sigma^2},
	\label{eq:richardson}
\ee
where $N$ is the Brunt-V\"{a}is\"{a}l\"{a} frequency
\be
	N^2 = \frac{g}{H}\frac{\chi_T}{\chi_\rho}
		\left[\nabla_{\rm ad}-\lp\frac{d\ln T}{d\ln P}\rp_*\right],
\ee
where $\chi_Q=\partial\ln P/\partial\ln Q$, with all other intensive variables set constant, $\nabla_{\rm ad}=(\partial\ln T/\partial\ln P)_{\rm ad}$ is the adiabatic temperature gradient, and the star refers to derivatives of the envelope's profile. This is just the thermal contribution to the buoyancy, and it is estimated to be \citep{Bildsten98}
\be
	N \approx \lp\frac{3}{20}\frac{g}{H}\rp^{1/2}
	=2.6\temp^{-1/2}\,{\rm s^{-1}}.
\ee
Note that for the scalings presented in this section I focus on the thermodynamic properties of the surface layers and omit the scalings with the WD mass, radius, and composition for simplicity. These scalings all assume $M=1\,M_\odot$, $R=5\times10^8\,{\rm cm}$, and a helium-rich composition, but I consider changes from these WD properties in the numerical investigation later.  Linear analysis shows that Kelvin-Helmholtz instability occurs when $Ri<1/4$, which develops into strong turbulence that readily transports angular momentum.

The condition for instability of $Ri<1/4$ assumes that thermal diffusion can be ignored for the unstable fluid perturbations, in other words, that the perturbations are adiabatic. Fluid perturbations with a characteristic size $L$ and speed $V$ become non-adiabatic when the timescale for thermal diffusion, $L^2/K$, where $K$ is the thermal diffusivity, is less than the timescale of the perturbation, $L/V$. The ratio of these two timescales is the P\'{e}clet number, $Pe\equiv VL/K$ \citep{Townsend58}. The restoring force provided by thermal buoyancy is weakened when $Pe<1$, which requires the substitution of $N^2\rightarrow PeN^2$ and promotes instability. Thermal diffusion is most efficient at small lengthscales, which motivates setting $LV/\nu_k=Re_c$ \citep{Zahn92}, where $\nu_k$ is the kinematic viscosity and $Re_c$ is the critical Reynolds number for turbulence, which is of order 1000. This gives the P\'{e}clet number approximately related to the Prandtl number, $Pr$, by $Pe\approx Re_cPr$. The turbulent perturbations are thus non-adiabatic when \citep{Zahn92}
\be
	K>\nu_k Re_c.
	\label{eq:secularcondition}
\ee
In the non-degenerate surface layers the kinematic viscosity is dominated by ions, and has a value of \citep{Spitzer62}
\be
	\nu_k =1.4\times10^{-2}\rho_5^{-1}T_8^{5/2}\,{\rm cm^2\ s^{-1}},
	\label{eq:ionviscosity}
\ee
where $\rho_5\equiv\rho/10^5\ {\rm g\ cm^{-3}}$, and I assume a Coulomb logarithm of $\ln\Lambda=20$. Setting $K =16\sigma_{\rm SB}T^3/(3c_p\kappa\rho^2)$, where $\sigma_{\rm SB}$ the Stefan-Boltzmann constant, $c_p$ the specific heat, and $\kappa$ the opacity, the thermal diffusivity is
\be
	K = 980\kappa_{0.2}^{-1}\dens^{-2}\temp^3\, {\rm cm^2\ s^{-1}},
	\label{eq:thermal}
\ee
where I approximate $c_p=5k_{\rm B}/2\mu m_p$ and scale the opacity to $\kappa_{0.2}\equiv\kappa/0.2\ {\rm cm^{2}\ g^{-1}}$, appropriate for electron scattering in hydrogen-deficient material. Substituting Equations (\ref{eq:ionviscosity}) and (\ref{eq:thermal}) into Equation (\ref{eq:secularcondition}), I find that the perturbations are non-adiabatic at depths of \mbox{$\rho\lesssim7\times10^6\,{\rm g\ cm^{-3}}\ T_8^{1/2}$}. The new ``secular'' Richardson number associated with this limit is,
\be
	Ri_s\equiv \frac{\nu_kRe_c}{K}\frac{N^2}{\sigma^2}.
	\label{eq:secular}
\ee
When $Ri_s<1/4$, the so-called ``secular shear instability'' arises.

   The competing effects of accretion increasing $\sigma$ versus turbulence developing when $Ri_s<1/4$ (and decreasing $\sigma$) drive the surface layers toward marginally satisfying $Ri_s=1/4$ (assuming for the moment that the sole viscous mechanism is the Kelvin-Helmholtz instability). This expectation is borne out in the white dwarf studies of \citet{Yoon04}. Thus I estimate the $q$ due to this mechanism. Substituting $Ri_s=1/4$ into Equation (\ref{eq:secular}), and assuming $Re_c=1000$, the shear rate is
\be
	\sigma_{\rm KH}=2.0\kappa_{0.2}^{1/2}\dens^{1/2}\temp^{-3/4}\,{\rm s^{-1}}.
	\label{eq:sigmakh}
\ee
A shear rate this large is similar to the Keplerian frequency given by Equation (\ref{eq:omegak}), but as I show next, other fluid instabilities limit the shear before it can reach such a large value.

\subsection{Baroclinic Instability}

   The baroclinic instability arises because surfaces of constant pressure and density no longer coincide if hydrostatic balance is to be maintained when differential rotation is present. In such a configuration, fluid perturbations along nearly horizontal directions are unstable, though with a sufficient radial component to allow mixing of angular momentum and material. The instability can roughly be broken into two limits, depending on a critical baroclinic Richardson number \citep{Fujimoto87},
\be
	Ri_{\rm BC} \equiv 4\lp\frac{R}{H}\rp^2\lp\frac{\Omega}{N}\rp^2
	= 10.7\temp^{-1}\spin^2,
	\label{eq:ri_bc}
\ee
where $\Omega_{0.1}=\Omega/0.1\,{\rm s^{-1}}$. When $Ri>Ri_{\rm BC}$, Coriolis effects limit the horizontal scale of perturbations. This results in two parameterizations for viscosity estimated from linear theory \citep{Fujimoto93},
\be
       \nu_{\rm BC} = \left\{
              \begin{array}{ccc}
        \displaystyle \frac{\alpha_{\rm BC}}{3}\frac{1}{Ri^{1/2}}H^2\Omega,
                &\hspace{0.2cm}&Ri\le Ri_{\rm BC},\\
         && \\
        \displaystyle \frac{\alpha_{\rm BC}}{3}\frac{Ri_{\rm BC}}{Ri^{3/2}}H^2\Omega,
                &\hspace{0.2cm}&Ri>Ri_{\rm BC},
              \end{array}
       \right.
       \label{eq:baroclinic}
\ee
where I include a dimensionless factor $\alpha_{\rm BC}$, to account for uncertainty in how linear theory relates to the saturated amplitudes of the instability. In general, I find $\alpha_{\rm BC}\lesssim1$ is required for significant mixing, which I explain in more detail below.

   By substituting $\nu_{\rm BC}$ into Equation (\ref{eq:angularmomentum2}), I solve for the shearing profile. Due to the relatively low $Ri_{\rm BC}$ for WDs, see Equation (\ref{eq:ri_bc}), it is almost always the case that $Ri>Ri_{\rm BC}$. Thus, using the second of the two viscosity prescriptions I find
\be
	Ri =3.6\times10^4\alpha_{\rm BC}^{1/2}\dens^{1/2}\temp^{1/4}\Omega_{0.1}^{3/2}\dot{M}_{-6}^{-1/2},
\ee
where $\dot{M}_{-6}=\dot{M}/10^{-6}\,M_\odot\,{\rm yr^{-1}}$. This demonstrates directly that $Ri\gg Ri_{\rm BC}$ for accreting WDs. The shear is
\be
	\sigma_{\rm BC} =
	1.4\times10^{-2}
	\alpha_{\rm BC}^{-1/4}\dens^{-1/4}\temp^{-5/8}\spin^{-3/4}\dot{M}_{-6}^{1/4}\,{\rm s^{-1}},
	\nonumber
	\\
	\label{eq:sigma_bc}
\ee
This demonstrates a couple of key features of the shear. First, it is higher for large accretion rates, which makes perfect sense because angular momentum is being added more quickly to the star. Second, it is smaller for higher spin rates. This is because the baroclinic instability becomes stronger when the WD is spinning faster, which leads to a smoothing out of the shearing. Also note that generally $\sigma_{\rm BC}\ll \sigma_{\rm KH}$, so that the baroclinic instability triggers before the Kelvin-Helmholtz instability. This prevents the shear rate from ever becoming large enough for the Kelvin-Helmholtz instability to operate at depths of $\rho\gtrsim100\ {\rm g\ cm^{-3}}$, i.e., at depths critical for unstable helium ignition.

\subsection{Viscous Heating}
\label{sec:heating}

The shear and dissipation by viscosity also leads to viscous heating within the accreted layer. The heating rate per unit mass is
\be
	\epsilon = \nu \sigma^2/2,
\ee
which for the shearing estimates given above results in
\be
	\epsilon_{\rm BC} = 6.9\times10^2\alpha_{\rm BC}^{-1/4}\rho_5^{-5/4}T_8^{-5/8}\Omega_{0.1}^{-3/4}\dot{M}_{-6}^{5/4}\,{\rm erg\,g^{-1}\,s^{-1}}.
	\nonumber
	\\
\ee
To put this in context, it is helpful to multiply this result by the accretion timescale $t_{\rm acc}$ to roughly give the total energy per nucleon accreted
\be
	E_{\rm BC} \approx 0.04\alpha_{\rm BC}^{-1/4}\rho_5^{-1/4}T_8^{3/8}\Omega_{0.1}^{-3/4}\dot{M}_{-6}^{1/4}\,{\rm keV\,nucl^{-1}}.
	\nonumber
	\\
\ee
In comparison, the thermal energy at the base of the accreted layer is about $\approx10\,{\rm keV\,nucl^{-1}}$. This means that viscous heating is usually not important, but can be for especially low values of $\alpha_{\rm BC}$ and $\Omega$ \citep[as can be see in discussions by][but in the context of shear instabilities]{Yoon04b}. For the remainder of this work I ignore viscous heating so as to focus on the impact of mixing, but heating should be included in a fuller, more self-consistent calculation.

\subsection{Turbulent Mixing Estimates}
\label{sec:mixing}

The turbulence that transports angular momentum also mixes material, albeit with much less efficiency because it requires more work to exchange fluid elements than to just exert stresses. Energy arguments give a mixing diffusivity that is related to the viscosity by \citep{Piro07}
\be
	D \approx \nu/Ri,
\ee
where there is a correction of order unity in this relation (often referred to as the flux Richardson number $Rf$), which I ignore for simplicity. I already have one free parameter in $\alpha_{\rm BC}$ to adjust the strength of the turbulence (although this does not allow me to vary the angular momentum and material mixing independently).

For this mixing diffusivity, I can now evaluate the mixing time as a function of depth using Equation (\ref{eq:t_mix}), resulting in
\be
	t_{\rm mix} = 2.1\times10^4 \alpha_{\rm BC}^{1/4}\rho_5^{5/4}T_8^{13/8}\Omega_{0.1}^{3/4}\dot{M}_{-6}^{-5/4}\,{\rm yr}.
\ee
In comparison, the accretion time down to a similar depth is
\be
	t_{\rm acc} = 4\pi R^2H\rho/\dot{M} = 1.9\times10^3\rho_5T_8\dot{M}_{-6}^{-1}\,{\rm yr}.
\ee
This shows that $t_{\rm mix}$ and $t_{\rm acc}$ are somewhat comparable for reasonable parameters, motivating that mixing is potentially important. Furthermore, since $t_{\rm mix}/t_{\rm acc}\propto \dot{M}^{-1/4}$, the mixing gets stronger as the accretion rate increases. Conversely, the scalings with density argue that mixing is only effective in the layers sufficiently shallow that $t_{\rm mix}<t_{\rm acc}$ or when
\be
	\rho < 5.9 \alpha_{\rm BC}^{-1} T_8^{-5/32}\Omega_{0.1}^{-3}\dot{M}_{-6}\,{\rm g\,cm^{-3}}.
	\label{eq:rho_mix}
\ee
This relatively small number for $\rho$ means that both the viscosity parameter must be  small $\alpha_{\rm BC}\sim10^{-2}$ and the spin must be relatively small $\Omega\sim10^{-2}\Omega_{\rm K}$ for mixing to happen. This motivates the values I use for these parameters for the numerical calculations in the next section. In particular, the large exponent of $\Omega^{-3}$ shows that the spin plays an especially important role in determining whether mixing is occurring. Also, note that the mixing I find here is never as large as what \citet{Yoon04c} calculate when focusing on the secular shear instability. This means that I also do not find solutions where the helium is mixed sufficiently enough to prevent unstable shell ignition.


\section{Numerical Calculations}
\label{sec:numerical}

I next explore the mixing in more detail with numerical models. For purposes of the present study, the profile for these conditions can be estimated with a constant flux atmosphere model, which extends down to the depth helium burning ignited unstably. This condition is given by the relation
\be
	\frac{d\epsilon_{\rm 3\alpha}}{dT} = \frac{d\epsilon_{\rm cool}}{dT},
	\label{eq:stability}
\ee
where $\epsilon_{\rm 3\alpha}$ is the triple-$\alpha$ heating rate \citep{Fushika87} and
\be
	\epsilon_{\rm cool} = \frac{4\sigma_{\rm SB}T^4}{3\kappa\Sigma^2}
\ee
is the radiative cooling rate. Both derivatives are taken at constant pressure. I solve for the temperature profile by solving the radiative diffusion equation,
\be
	F = \frac{16\sigma_{\rm SB}T^3}{3\kappa}\frac{dT}{d\Sigma},
\ee
where I set the flux to a value which results in the correct ignition depth for the given WD mass and accretion rate. Since I am ignoring viscous heating, I can use the same temperature profile whether or not turbulent mixing is included. The opacity is dominated by electron scattering at the relatively hot temperatures of the high accretion rates I consider.

When mixing is included, it is incorporated as described in Section \ref{sec:mixing framework}. Summarizing here, I let the WD accrete for a time $t_{\rm acc}$. I then solve for the mixing depth with an iterative process until the base of the layer correctly satisfies $t_{\rm mix}=t_{\rm acc}$. Once I know the depth of the mixing, I can compare the base conditions to Equation (\ref{eq:stability}), making sure to use the correctly found $Y_{\rm mix}$ in the stability criterion. If $d\epsilon_{\rm 3\alpha}/dT < d\epsilon_{\rm cool}/dT$, then the layer is stable and accretion can continue. Once $d\epsilon_{\rm 3\alpha}/dT =d\epsilon_{\rm cool}/dT$, then the layer unstably ignites, and I can record the amount of mixing in that given model.

\subsection{Example Accumulating Model}

To make this scheme more concrete, it is helpful to to consider a specific example of the time evolution of an envelope as it accretes toward unstable ignition. This example is for a $M=1\,M_\odot$ WD accreting at a rate $\dot{M}=10^{-6}\,M_\odot\,{\rm yr^{-1}}$. A reasonable ignition mass for such conditions is $M_{\rm ign}\approx 3.0\times10^{-3}\,M_\odot$ \citep{Iben89,Shen09}. For the mixing, I set the WD spin to be $\Omega=0.01\Omega_{\rm K}$ and use $\alpha_{\rm BC}=0.01$.

\begin{figure*}
\centering
\includegraphics[width=0.6\textwidth]{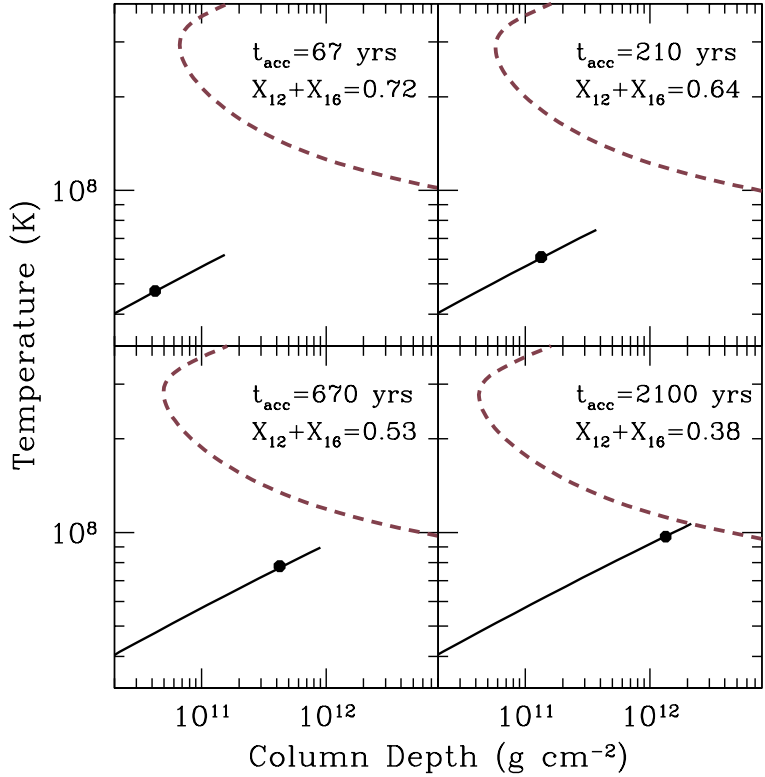}
\caption{Four snapshots in time as an accreted layer grows toward unstable ignition. For this example, $M=1\,M_\odot$, $R=5\times10^8\,{\rm cm}$, $\Omega=0.01\Omega_{\rm K}$, and $\alpha_{\rm BC}=0.01$. Each panel shows the envelope temperature $T$ versus column depth $\Sigma$ profile as a solid line. The dashed curves indicate where unstable ignition occurs, which is found using Equation (\ref{eq:stability}). The filled circles indicate the depth to which the envelope would extend if mixing was not included. The total mass fraction of C/O material mixed into the surface helium layer is denoted in each panel. Note that I denote the mass fraction variables as $X_{12}$ and $X_{16}$ even though these refer to the mixed fraction (e.g., I have removed the subscript ``mix'' for brevity).}
\label{fig:snapshots}
\epsscale{1.0}
\end{figure*}

Figure \ref{fig:snapshots} shows four different moments in time as the WD accretes and builds a surface layer of accreted material mixed with C/O from the WD surface. In each panel, the amount of time for which the WD has been accreting $t_{\rm acc}$ is denoted along with the total mass fraction of C/O mixed into the surface layer. The solid line shows the temperature profile down to the base of the layer, with the solid circle denoting the amount of material that was actually accreted (much like the horizontal dashed line in Figure \ref{fig:diagram}). The dashed curve is the stability criterion given by Equation (\ref{eq:stability}). In the bottom right panel the surface temperature profile now extends down to this ignition curve demonstrating that this model will now ignite.

A number of important features and trends caused by the mixing, that were described previously in my analytic exploration in Section \ref{sec:equations}, are now made more clear by this numerical example. These are as follows.
\begin{enumerate}
\item Mixing is strongest at earlier times and decreases the longer the accretion persists until ignition (as can be seen by the mixed C/O mass fraction decreasing with time). This because it gets more and more difficult to mix to larger depths.
\item Mixing causes unstable ignition of the surface layer in a larger mass of material for a given accretion rate than without mixing. In this case the mass of the surface layer is $3.4\times10^{-3}\,M_\odot$ at the moment of ignition after accreting $2.1\times10^{-3}\,M_\odot$ of helium. This is $\approx13\%$ more mass in the surface layer than the unmixed case.
\item Igniting deeper also decreases the time to accrete until ignition, in this case by $\approx30\%$ from $\approx3\times10^3\,{\rm yrs}$ down to $\approx2.1\times10^{3}\,{\rm yrs}$.
\end{enumerate}
Most importantly of all, this demonstrates that it is reasonable for ignition to occur in a fairly mixed environment. In this case, $Y_{\rm mix}= 0.62$ while the mass fraction of C/O is $0.38$. This hopefully motivates the investigation of similarly mixed conditions in future surface burning models \citep[e.g.,][]{Shen14b}.

\subsection{Trends with Accretion Rate and Spin}

\begin{figure}
\epsscale{1.2}
\plotone{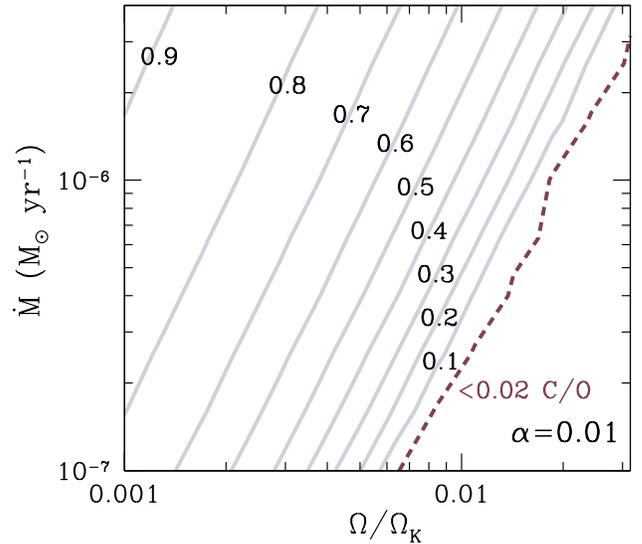}
\caption{Solid lines are contours of constant mass fraction of C/O in the mixed accreted layer as a function of the accretion WD spin $\Omega$ and the accretion rate $\dot{M}$. This calculation uses $M=1\,M_\odot$, $R=5\times10^8\,{\rm cm}$, and $\alpha_{\rm BC}=0.01$. These contours are spaced every 0.1 in mass fraction as labeled. The dashed curve shows the boundary at which to the right less than 0.02 of C/O is mixed up into the accreted material. The wiggles in the dashed line are numerical and do not represent a real physical effect.}
\label{fig:ymix}
\epsscale{1.0}
\end{figure}

To further illustrate the trends of the mixing, I explore the amount of mixing as a function of the accretion rate and WD spin. In Figure \ref{fig:ymix}, I use $M=1\,M_\odot$, $R=5\times10^8\,{\rm cm}$, and $\alpha_{\rm BC}=0.01$ and plot contours of the mass fraction of C/O. This shows there is quite a range of mixing, from nearly unmixed on the right side of the plot to fully 90\% C/O on the left side.

The general trend is that more mixing occurs at high accretion rates and low WD spin. The impact of a high accretion is intuitively clear, because if more angular momentum and material is added more quickly it is reasonable to expect more mixing. Less clear is why slower spin leads to more mixing. The first thing to note is that it is {\it not} because there is a larger shear between the WD and the accreted material; the total change in spin is basically the same whether $\Omega=0.001\Omega_{\rm K}$ or $0.01\Omega_{\rm K}$. It is instead because the viscosity scales with the spin as shown in Equation (\ref{eq:baroclinic}). As the WD spins higher, surfaces of constant pressure and density become more and more misaligned, which in turn drives the baroclinic instability and turbulent viscosity associated with it. A higher viscosity results in a smaller shear to transport the angular momentum, as shown in Equation (\ref{eq:angularmomentum2}), and less shear leads to less mixing. Conversely, for small spin and turbulent viscosity, the shear is large and thus the mixing with it. This is particularly striking in Figure \ref{fig:ymix}, where to the right of the dashed line there is quite little mixing.

\begin{figure}
\epsscale{1.2}
\plotone{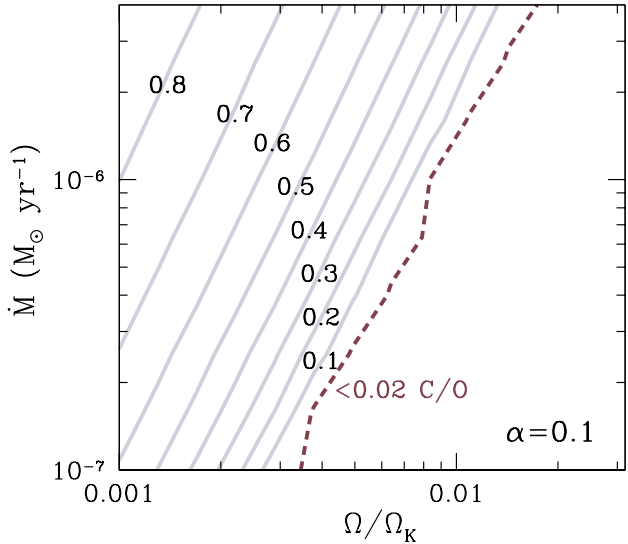}
\caption{The same as Figure \ref{fig:ymix}, but with $\alpha=0.1$.}
\label{fig:ymix2}
\epsscale{1.0}
\end{figure}

Since the analytic estimates showed a strong dependence on $\alpha_{\rm BC}$ in Equation (\ref{eq:rho_mix}), I present the mixed C/O mass fractions for $\alpha_{\rm BC}=0.1$ in Figure \ref{fig:ymix2}. This roughly moves all the curves to the left and up, showing how much more difficult it is to mix when the viscosity is higher. This is simply because there is less shear and turbulence in this case. For $\alpha_{\rm BC}\sim1$ there is little mixing except in the most extreme circumstances (of high accretion rate and low spin), and thus it is possible that turbulent mixing is negligible depending on what the real turbulent viscosity should be.

\begin{figure}
\epsscale{1.2}
\plotone{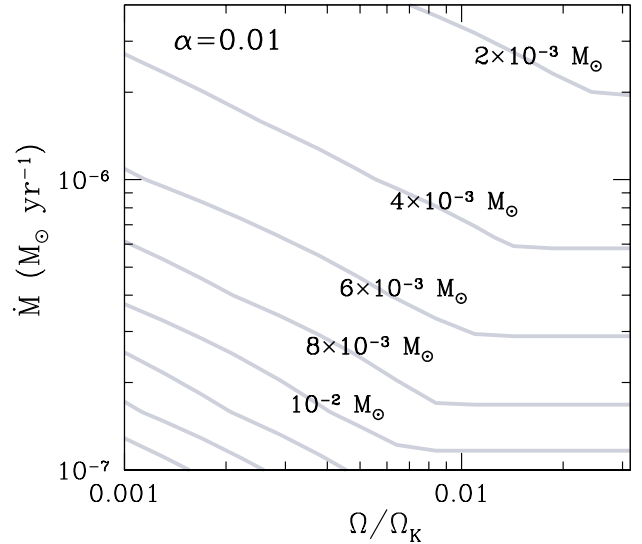}
\caption{The same as Figure \ref{fig:ymix}, but the contours are instead for constant ignition mass.  Each curve is spaced by $2\times10^{-3}\,M_\odot$, some of which are labeled.}
\label{fig:mign}
\epsscale{1.0}
\end{figure}

As mentioned in the example calculation above, the mixing also causes ignition deeper than without mixing. This is quantified further in Figure \ref{fig:mign}, where I plot contours of constant ignition mass. This shows that at fixed accretion rate, the ignition mass goes up for lower spin (i.e., moving to the left). At fixed spin, the ignition mass goes down for higher accretion rate just as in the non-mixed case, but just not by as much.

\subsection{Trends with White Dwarf Mass}

\begin{figure}
\epsscale{1.2}
\plotone{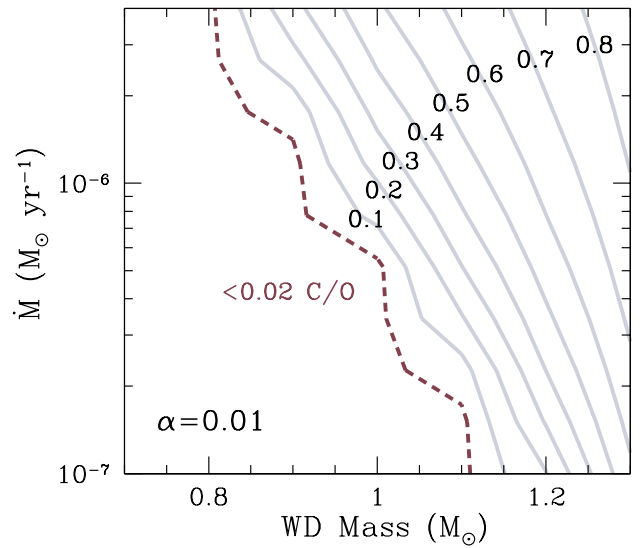}
\caption{Similar to Figure \ref{fig:ymix}, but now mixed C/O mass fraction as a function of the WD mass.}
\label{fig:ymix_mass}
\epsscale{1.0}
\end{figure}

The strength of the mixing should also depend on the mass and radius of the WD. To explore this, I repeat my analysis of for the amount of mixing which is summarized in Figure \ref{fig:ymix_mass}. This shows that there is in fact a strong dependence on mass. This can be understood from two main factors. First, the Keplerian spin of the incoming material is larger with larger mass, which means a larger amount of angular momentum needs to be carried through the star. Second, the larger surface gravity in turn results in a smaller scale height $H$ in the surface layers. This weakened the turbulent viscosity, leading to a larger shear and more mixing.

\section{Discussion and Conclusions}
\label{sec:conclusions}

I have explored mixing in the surface layers of helium-accreting WDs via hydrodynamic instabilities. These models highlight a number of important trends for mixing, which are as follows.
\begin{enumerate}
\item Mixing is greatest at low spin $\Omega$, high accretion rate $\dot{M}$, and high WD mass $M$.
\item Mixing is stronger for low $\alpha_{\rm BC}$, and for $\alpha_{\rm BC}\sim1$ mixing is negligible.
\item Mixing causes unstable ignition of the surface layer to occur sooner and involve more mass.
\item Depending on the parameters, ignition of the surface layer can occur in highly mixed conditions with C/O mass fractions sometimes $\gtrsim0.5$.
\end{enumerate}
These results will hopefully motivate studies that follow the evolution of the surface layers after ignition occurs \citep[e.g.,][]{Shen14b}, but including strong mixing and larger ignition depth. In particular, it will be important to understand how the nucleosynthetic products are different. Historically, the double detonation scenario has had trouble with a high abundance of iron-peak elements near the surface of the star which causes the colors of the resulting SN Ia to not match those observed \citep{Woosley11,Sim12}, but a large amount of mixing may alleviate this problem. The different composition may also imprint itself on the properties of the rising light SN Ia light curve \citep{Piro12,Piro13,Piro14}. Furthermore, mixing tends to make more calcium-rich burning products. In particular, \citet{Waldman11} find that $\approx30\%$ carbon mixed into the accreted helium may help explain the calcium-rich composition in a subset of rapid transients \citep{Kasliwal12}.

The study presented here is mainly focused on highlighting the main properties of turbulent mixing, and thus there are many additional features that still need to be investigated in more detail. Most notably, the ignition conditions and envelope profiles were not calculated from first principles. Instead the known ignition conditions from the literature were used for the non-mixed cases, and then these were altered in a reasonable way to infer the changes from mixing. This can be improved by implementing mixing into accreting WD models in a stellar evolution code like \texttt{MESA} \citep{Paxton13}, so that one can follow the evolution of the WD up to the point of ignition. The mixing would likely have to be implemented as a diffusive scheme \citep[e.g.,][]{Heger00}. Although, as I argued earlier, from a physical perspective this is not necessarily a more accurate description of what occurs during turbulent mixing. Nevertheless, comparing this scheme to the one I use here is also an important exercise (such a comparison has been done in the case of accreting neutron stars by \citealp{Piro07} and \citealp{Keek09}, which shows at least qualitatively similar results). These studies could also explore the impact of other burning chains up until the point of ignition (besides the triple-$\alpha$ I focus on here) and viscous heating (see the discussion in Section \ref{sec:heating}).

Another potentially important extension of this work is hydrogen-rich accretion and the subsequent classical novae (CNe) that result when mixing is included. Observation indicate that some ejecta of CNe are enriched by $\approx30\%$ in C, N, O, and Ne \citep{Livio94,Gehrz98}. Furthermore, a similar amount of enrichment has been argued on theoretical grounds to explain the energetics of the ejecta \citep{Starrfield72,Truran82} as well give the best match with one-dimensional models of the ignition \citep{Hernanz96,Starrfield98}. The method for enriching CNe ejecta with heavier elements remains an outstanding theoretical question. If the enrichment is indeed important for ignition, then it must take place prior to or during the early stages of the thermonuclear runaway. Previous theoretical investigations include diffusion of hydrogen at low accretion rates \citep[$\lesssim10^{-10}\ M_\odot\ {\rm yr^{-1}}$,][]{Prialnik84,Kovetz85}. Others have looked at convective overshoot \citep{Denissenkov14}, motivated by the work of \citet{Casanova11}. Turbulent mixing from accretion, as presented here, should also be explored as a possible solution, and may be naturally stronger if accreting WDs are spinning slowly \citep{Livio98} or on ONe WDs \citep{Mason11} because of their higher masses.

\acknowledgements
I thank Ken Shen for useful exchanges on the double detonation scenario and surface burning on WDs, and I thank Ashley Ruiter and Stuart Sim for feedback on a previous draft. I also thank Mark Phillips for his generosity in arranging my visit to Las Campanas Observatory where much of this work was completed.


\bibliographystyle{apj}
\bibliography{ms}

\begin{thebibliography}{}
\expandafter\ifx\csname natexlab\endcsname\relax\def\natexlab#1{#1}\fi

\bibitem[{{Bildsten}(1998)}]{Bildsten98}
{Bildsten}, L. 1998, in NATO Advanced Science Institutes (ASI) Series C, ed.
  R.~{Buccheri}, J.~{van Paradijs}, \& A.~{Alpar}, Vol. 515, 419

\bibitem[{{Bildsten} {et~al.}(2007){Bildsten}, {Shen}, {Weinberg}, \&
  {Nelemans}}]{Bildsten07}
{Bildsten}, L., {Shen}, K.~J., {Weinberg}, N.~N., \& {Nelemans}, G. 2007,
  \apjl, 662, L95

\bibitem[{{Blondin} {et~al.}(2012){Blondin}, {Matheson}, {Kirshner}, {Mandel},
  {Berlind}, {Calkins}, {Challis}, {Garnavich}, {Jha}, {Modjaz}, {Riess}, \&
  {Schmidt}}]{Blondin12}
{Blondin}, S., {Matheson}, T., {Kirshner}, R.~P., {et~al.} 2012, \aj, 143, 126

\bibitem[{{Bloom} {et~al.}(2012){Bloom}, {Kasen}, {Shen}, {Nugent}, {Butler},
  {Graham}, {Howell}, {Kolb}, {Holmes}, {Haswell}, {Burwitz}, {Rodriguez}, \&
  {Sullivan}}]{Bloom12}
{Bloom}, J.~S., {Kasen}, D., {Shen}, K.~J., {et~al.} 2012, \apjl, 744, L17

\bibitem[{{Cantiello} {et~al.}(2014){Cantiello}, {Mankovich}, {Bildsten},
  {Christensen-Dalsgaard}, \& {Paxton}}]{Cantiello14}
{Cantiello}, M., {Mankovich}, C., {Bildsten}, L., {Christensen-Dalsgaard}, J.,
  \& {Paxton}, B. 2014, \apj, 788, 93

\bibitem[{{Casanova} {et~al.}(2011){Casanova}, {Jos{\'e}},
  {Garc{\'{\i}}a-Berro}, {Shore}, \& {Calder}}]{Casanova11}
{Casanova}, J., {Jos{\'e}}, J., {Garc{\'{\i}}a-Berro}, E., {Shore}, S.~N., \&
  {Calder}, A.~C. 2011, \nat, 478, 490

\bibitem[{{Childress} {et~al.}(2014){Childress}, {Filippenko}, {Ganeshalingam},
  \& {Schmidt}}]{Childress14}
{Childress}, M.~J., {Filippenko}, A.~V., {Ganeshalingam}, M., \& {Schmidt},
  B.~P. 2014, \mnras, 437, 338

\bibitem[{{Denissenkov} {et~al.}(2014){Denissenkov}, {Truran}, {Pignatari},
  {Trappitsch}, {Ritter}, {Herwig}, {Battino}, {Setoodehnia}, \&
  {Paxton}}]{Denissenkov14}
{Denissenkov}, P.~A., {Truran}, J.~W., {Pignatari}, M., {et~al.} 2014, \mnras,
  442, 2058

\bibitem[{{Fink} {et~al.}(2007){Fink}, {Hillebrandt}, \& {R{\"o}pke}}]{Fink07}
{Fink}, M., {Hillebrandt}, W., \& {R{\"o}pke}, F.~K. 2007, \aap, 476, 1133

\bibitem[{{Fink} {et~al.}(2010){Fink}, {R{\"o}pke}, {Hillebrandt},
  {Seitenzahl}, {Sim}, \& {Kromer}}]{Fink10}
{Fink}, M., {R{\"o}pke}, F.~K., {Hillebrandt}, W., {et~al.} 2010, \aap, 514,
  A53

\bibitem[{{Foley} {et~al.}(2013){Foley}, {Challis}, {Chornock},
  {Ganeshalingam}, {Li}, {Marion}, {Morrell}, {Pignata}, {Stritzinger},
  {Silverman}, {Wang}, {Anderson}, {Filippenko}, {Freedman}, {Hamuy}, {Jha},
  {Kirshner}, {McCully}, {Persson}, {Phillips}, {Reichart}, \&
  {Soderberg}}]{Foley13}
{Foley}, R.~J., {Challis}, P.~J., {Chornock}, R., {et~al.} 2013, \apj, 767, 57

\bibitem[{{Fujimoto}(1987)}]{Fujimoto87}
{Fujimoto}, M.~Y. 1987, \aap, 176, 53

\bibitem[{{Fujimoto}(1993)}]{Fujimoto93}
---. 1993, \apj, 419, 768

\bibitem[{{Fushiki} \& {Lamb}(1987)}]{Fushika87}
{Fushiki}, I., \& {Lamb}, D.~Q. 1987, \apj, 317, 368

\bibitem[{{Gehrz} {et~al.}(1998){Gehrz}, {Truran}, {Williams}, \&
  {Starrfield}}]{Gehrz98}
{Gehrz}, R.~D., {Truran}, J.~W., {Williams}, R.~E., \& {Starrfield}, S. 1998,
  \pasp, 110, 3

\bibitem[{{Hancock} {et~al.}(2011){Hancock}, {Gaensler}, \&
  {Murphy}}]{Hancock11}
{Hancock}, P.~J., {Gaensler}, B.~M., \& {Murphy}, T. 2011, \apjl, 735, L35

\bibitem[{{Hayden} {et~al.}(2010){Hayden}, {Garnavich}, {Kasen}, {Dilday},
  {Frieman}, {Jha}, {Lampeitl}, {Nichol}, {Sako}, {Schneider}, {Smith},
  {Sollerman}, \& {Wheeler}}]{Hayden10}
{Hayden}, B.~T., {Garnavich}, P.~M., {Kasen}, D., {et~al.} 2010, \apj, 722,
  1691

\bibitem[{{Heger} {et~al.}(2000){Heger}, {Langer}, \& {Woosley}}]{Heger00}
{Heger}, A., {Langer}, N., \& {Woosley}, S.~E. 2000, \apj, 528, 368

\bibitem[{{Hernanz} {et~al.}(1996){Hernanz}, {Jose}, {Coc}, \&
  {Isern}}]{Hernanz96}
{Hernanz}, M., {Jose}, J., {Coc}, A., \& {Isern}, J. 1996, \apjl, 465, L27

\bibitem[{{Holcomb} {et~al.}(2013){Holcomb}, {Guillochon}, {De Colle}, \&
  {Ramirez-Ruiz}}]{Holocomb13}
{Holcomb}, C., {Guillochon}, J., {De Colle}, F., \& {Ramirez-Ruiz}, E. 2013,
  \apj, 771, 14

\bibitem[{{Horesh} {et~al.}(2012){Horesh}, {Kulkarni}, {Fox}, {Carpenter},
  {Kasliwal}, {Ofek}, {Quimby}, {Gal-Yam}, {Cenko}, {de Bruyn}, {Kamble},
  {Wijers}, {van der Horst}, {Kouveliotou}, {Podsiadlowski}, {Sullivan},
  {Maguire}, {Howell}, {Nugent}, {Gehrels}, {Law}, {Poznanski}, \&
  {Shara}}]{Horesh12}
{Horesh}, A., {Kulkarni}, S.~R., {Fox}, D.~B., {et~al.} 2012, \apj, 746, 21

\bibitem[{{Iben} \& {Tutukov}(1984)}]{Iben84}
{Iben}, Jr., I., \& {Tutukov}, A.~V. 1984, \apjs, 54, 335

\bibitem[{{Iben} \& {Tutukov}(1989)}]{Iben89}
---. 1989, \apj, 342, 430

\bibitem[{{Inserra} {et~al.}(2015){Inserra}, {Sim}, {Wyrzykowski}, {Smartt},
  {Fraser}, {Nicholl}, {Shen}, {Jerkstrand}, {Gal-Yam}, {Howell}, {Maguire},
  {Mazzali}, {Valenti}, {Taubenberger}, {Benitez-Herrera}, {Bersier},
  {Blagorodnova}, {Campbell}, {Chen}, {Elias-Rosa}, {Hillebrandt},
  {Kostrzewa-Rutkowska}, {Koz{\l}owski}, {Kromer}, {Lyman}, {Polshaw},
  {R{\"o}pke}, {Ruiter}, {Smith}, {Spiro}, {Sullivan}, {Yaron}, {Young}, \&
  {Yuan}}]{Inserra15}
{Inserra}, C., {Sim}, S.~A., {Wyrzykowski}, L., {et~al.} 2015, \apjl, 799, L2

\bibitem[{{Kasen}(2010)}]{Kasen10}
{Kasen}, D. 2010, \apj, 708, 1025

\bibitem[{{Kasliwal} {et~al.}(2010){Kasliwal}, {Kulkarni}, {Gal-Yam}, {Yaron},
  {Quimby}, {Ofek}, {Nugent}, {Poznanski}, {Jacobsen}, {Sternberg}, {Arcavi},
  {Howell}, {Sullivan}, {Rich}, {Burke}, {Brimacombe}, {Milisavljevic},
  {Fesen}, {Bildsten}, {Shen}, {Cenko}, {Bloom}, {Hsiao}, {Law}, {Gehrels},
  {Immler}, {Dekany}, {Rahmer}, {Hale}, {Smith}, {Zolkower}, {Velur},
  {Walters}, {Henning}, {Bui}, \& {McKenna}}]{Kasliwal10}
{Kasliwal}, M.~M., {Kulkarni}, S.~R., {Gal-Yam}, A., {et~al.} 2010, \apjl, 723,
  L98

\bibitem[{{Kasliwal} {et~al.}(2012){Kasliwal}, {Kulkarni}, {Gal-Yam}, {Nugent},
  {Sullivan}, {Bildsten}, {Yaron}, {Perets}, {Arcavi}, {Ben-Ami}, {Bhalerao},
  {Bloom}, {Cenko}, {Filippenko}, {Frail}, {Ganeshalingam}, {Horesh}, {Howell},
  {Law}, {Leonard}, {Li}, {Ofek}, {Polishook}, {Poznanski}, {Quimby},
  {Silverman}, {Sternberg}, \& {Xu}}]{Kasliwal12}
---. 2012, \apj, 755, 161

\bibitem[{{Keek} {et~al.}(2009){Keek}, {Langer}, \& {in't Zand}}]{Keek09}
{Keek}, L., {Langer}, N., \& {in't Zand}, J.~J.~M. 2009, \aap, 502, 871

\bibitem[{{Kovetz} \& {Prialnik}(1985)}]{Kovetz85}
{Kovetz}, A., \& {Prialnik}, D. 1985, \apj, 291, 812

\bibitem[{{Kromer} {et~al.}(2010){Kromer}, {Sim}, {Fink}, {R{\"o}pke},
  {Seitenzahl}, \& {Hillebrandt}}]{Kromer10}
{Kromer}, M., {Sim}, S.~A., {Fink}, M., {et~al.} 2010, \apj, 719, 1067

\bibitem[{{Leonard}(2007)}]{Leonard07}
{Leonard}, D.~C. 2007, \apj, 670, 1275

\bibitem[{{Li} {et~al.}(2011){Li}, {Bloom}, {Podsiadlowski}, {Miller}, {Cenko},
  {Jha}, {Sullivan}, {Howell}, {Nugent}, {Butler}, {Ofek}, {Kasliwal},
  {Richards}, {Stockton}, {Shih}, {Bildsten}, {Shara}, {Bibby}, {Filippenko},
  {Ganeshalingam}, {Silverman}, {Kulkarni}, {Law}, {Poznanski}, {Quimby},
  {McCully}, {Patel}, {Maguire}, \& {Shen}}]{Li11}
{Li}, W., {Bloom}, J.~S., {Podsiadlowski}, P., {et~al.} 2011, \nat, 480, 348

\bibitem[{{Livio} \& {Pringle}(1998)}]{Livio98}
{Livio}, M., \& {Pringle}, J.~E. 1998, \apj, 505, 339

\bibitem[{{Livio} \& {Truran}(1994)}]{Livio94}
{Livio}, M., \& {Truran}, J.~W. 1994, \apj, 425, 797

\bibitem[{{Livne} \& {Arnett}(1995)}]{Livne95}
{Livne}, E., \& {Arnett}, D. 1995, \apj, 452, 62

\bibitem[{{Maguire} {et~al.}(2014){Maguire}, {Sullivan}, {Pan}, {Gal-Yam},
  {Hook}, {Howell}, {Nugent}, {Mazzali}, {Chotard}, {Clubb}, {Filippenko},
  {Kasliwal}, {Kandrashoff}, {Poznanski}, {Saunders}, {Silverman}, {Walker}, \&
  {Xu}}]{Maguire14}
{Maguire}, K., {Sullivan}, M., {Pan}, Y.-C., {et~al.} 2014, \mnras, 444, 3258

\bibitem[{{Mason}(2011)}]{Mason11}
{Mason}, E. 2011, \aap, 532, L11

\bibitem[{{Mazzali} {et~al.}(2005){Mazzali}, {Benetti}, {Altavilla}, {Blanc},
  {Cappellaro}, {Elias-Rosa}, {Garavini}, {Goobar}, {Harutyunyan}, {Kotak},
  {Leibundgut}, {Lundqvist}, {Mattila}, {Mendez}, {Nobili}, {Pain},
  {Pastorello}, {Patat}, {Pignata}, {Podsiadlowski}, {Ruiz-Lapuente}, {Salvo},
  {Schmidt}, {Sollerman}, {Stanishev}, {Stehle}, {Tout}, {Turatto}, \&
  {Hillebrandt}}]{Mazzali05}
{Mazzali}, P.~A., {Benetti}, S., {Altavilla}, G., {et~al.} 2005, \apjl, 623,
  L37

\bibitem[{{Moll} \& {Woosley}(2013)}]{Moll13}
{Moll}, R., \& {Woosley}, S.~E. 2013, \apj, 774, 137

\bibitem[{{Paxton} {et~al.}(2013){Paxton}, {Cantiello}, {Arras}, {Bildsten},
  {Brown}, {Dotter}, {Mankovich}, {Montgomery}, {Stello}, {Timmes}, \&
  {Townsend}}]{Paxton13}
{Paxton}, B., {Cantiello}, M., {Arras}, P., {et~al.} 2013, \apjs, 208, 4

\bibitem[{{Perets} {et~al.}(2010){Perets}, {Gal-Yam}, {Mazzali}, {Arnett},
  {Kagan}, {Filippenko}, {Li}, {Arcavi}, {Cenko}, {Fox}, {Leonard}, {Moon},
  {Sand}, {Soderberg}, {Anderson}, {James}, {Foley}, {Ganeshalingam}, {Ofek},
  {Bildsten}, {Nelemans}, {Shen}, {Weinberg}, {Metzger}, {Piro}, {Quataert},
  {Kiewe}, \& {Poznanski}}]{Perets10}
{Perets}, H.~B., {Gal-Yam}, A., {Mazzali}, P.~A., {et~al.} 2010, \nat, 465, 322

\bibitem[{{Perlmutter} {et~al.}(1999){Perlmutter}, {Aldering}, {Goldhaber},
  {Knop}, {Nugent}, {Castro}, {Deustua}, {Fabbro}, {Goobar}, {Groom}, {Hook},
  {Kim}, {Kim}, {Lee}, {Nunes}, {Pain}, {Pennypacker}, {Quimby}, {Lidman},
  {Ellis}, {Irwin}, {McMahon}, {Ruiz-Lapuente}, {Walton}, {Schaefer}, {Boyle},
  {Filippenko}, {Matheson}, {Fruchter}, {Panagia}, {Newberg}, {Couch}, \&
  {Project}}]{Perlmutter99}
{Perlmutter}, S., {Aldering}, G., {Goldhaber}, G., {et~al.} 1999, \apj, 517,
  565

\bibitem[{{Phillips}(1993)}]{Phillips93}
{Phillips}, M.~M. 1993, \apjl, 413, L105

\bibitem[{{Piro}(2008)}]{Piro08}
{Piro}, A.~L. 2008, \apj, 679, 616

\bibitem[{{Piro}(2012)}]{Piro12}
---. 2012, \apj, 759, 83

\bibitem[{{Piro} \& {Bildsten}(2004)}]{Piro04}
{Piro}, A.~L., \& {Bildsten}, L. 2004, \apj, 610, 977

\bibitem[{{Piro} \& {Bildsten}(2007)}]{Piro07}
---. 2007, \apj, 663, 1252

\bibitem[{{Piro} {et~al.}(2010){Piro}, {Chang}, \& {Weinberg}}]{Piro10}
{Piro}, A.~L., {Chang}, P., \& {Weinberg}, N.~N. 2010, \apj, 708, 598

\bibitem[{{Piro} \& {Nakar}(2013)}]{Piro13}
{Piro}, A.~L., \& {Nakar}, E. 2013, \apj, 769, 67

\bibitem[{{Piro} \& {Nakar}(2014)}]{Piro14}
---. 2014, \apj, 784, 85

\bibitem[{{Prialnik} \& {Kovetz}(1984)}]{Prialnik84}
{Prialnik}, D., \& {Kovetz}, A. 1984, \apj, 281, 367

\bibitem[{{Riess} {et~al.}(1998){Riess}, {Filippenko}, {Challis},
  {Clocchiatti}, {Diercks}, {Garnavich}, {Gilliland}, {Hogan}, {Jha},
  {Kirshner}, {Leibundgut}, {Phillips}, {Reiss}, {Schmidt}, {Schommer},
  {Smith}, {Spyromilio}, {Stubbs}, {Suntzeff}, \& {Tonry}}]{Riess98}
{Riess}, A.~G., {Filippenko}, A.~V., {Challis}, P., {et~al.} 1998, \aj, 116,
  1009

\bibitem[{{Ruiter} {et~al.}(2011){Ruiter}, {Belczynski}, {Sim}, {Hillebrandt},
  {Fryer}, {Fink}, \& {Kromer}}]{Ruiter11}
{Ruiter}, A.~J., {Belczynski}, K., {Sim}, S.~A., {et~al.} 2011, \mnras, 417,
  408

\bibitem[{{Ruiter} {et~al.}(2014){Ruiter}, {Belczynski}, {Sim}, {Seitenzahl},
  \& {Kwiatkowski}}]{Ruiter14}
{Ruiter}, A.~J., {Belczynski}, K., {Sim}, S.~A., {Seitenzahl}, I.~R., \&
  {Kwiatkowski}, D. 2014, \mnras, 440, L101

\bibitem[{{Scalzo} {et~al.}(2014){Scalzo}, {Ruiter}, \& {Sim}}]{Scalzo14}
{Scalzo}, R.~A., {Ruiter}, A.~J., \& {Sim}, S.~A. 2014, \mnras, 445, 2535

\bibitem[{{Schaefer} \& {Pagnotta}(2012)}]{Schaefer12}
{Schaefer}, B.~E., \& {Pagnotta}, A. 2012, \nat, 481, 164

\bibitem[{{Shappee} {et~al.}(2013){Shappee}, {Stanek}, {Pogge}, \&
  {Garnavich}}]{Shappee13}
{Shappee}, B.~J., {Stanek}, K.~Z., {Pogge}, R.~W., \& {Garnavich}, P.~M. 2013,
  \apjl, 762, L5

\bibitem[{{Shen} \& {Bildsten}(2009)}]{Shen09}
{Shen}, K.~J., \& {Bildsten}, L. 2009, \apj, 699, 1365

\bibitem[{{Shen} \& {Bildsten}(2014)}]{Shen14}
---. 2014, \apj, 785, 61

\bibitem[{{Shen} {et~al.}(2010){Shen}, {Kasen}, {Weinberg}, {Bildsten}, \&
  {Scannapieco}}]{Shen10}
{Shen}, K.~J., {Kasen}, D., {Weinberg}, N.~N., {Bildsten}, L., \&
  {Scannapieco}, E. 2010, \apj, 715, 767

\bibitem[{{Shen} \& {Moore}(2014)}]{Shen14b}
{Shen}, K.~J., \& {Moore}, K. 2014, ArXiv e-prints, arXiv:1409.3568

\bibitem[{{Sim} {et~al.}(2012){Sim}, {Fink}, {Kromer}, {R{\"o}pke}, {Ruiter},
  \& {Hillebrandt}}]{Sim12}
{Sim}, S.~A., {Fink}, M., {Kromer}, M., {et~al.} 2012, \mnras, 420, 3003

\bibitem[{{Sim} {et~al.}(2010){Sim}, {R{\"o}pke}, {Hillebrandt}, {Kromer},
  {Pakmor}, {Fink}, {Ruiter}, \& {Seitenzahl}}]{Sim10}
{Sim}, S.~A., {R{\"o}pke}, F.~K., {Hillebrandt}, W., {et~al.} 2010, \apjl, 714,
  L52

\bibitem[{{Spitzer}(1962)}]{Spitzer62}
{Spitzer}, L. 1962, {Physics of Fully Ionized Gases}

\bibitem[{{Spruit}(2002)}]{Spruit02}
{Spruit}, H.~C. 2002, \aap, 381, 923

\bibitem[{{Starrfield} {et~al.}(1972){Starrfield}, {Truran}, {Sparks}, \&
  {Kutter}}]{Starrfield72}
{Starrfield}, S., {Truran}, J.~W., {Sparks}, W.~M., \& {Kutter}, G.~S. 1972,
  \apj, 176, 169

\bibitem[{{Starrfield} {et~al.}(1998){Starrfield}, {Truran}, {Wiescher}, \&
  {Sparks}}]{Starrfield98}
{Starrfield}, S., {Truran}, J.~W., {Wiescher}, M.~C., \& {Sparks}, W.~M. 1998,
  \mnras, 296, 502

\bibitem[{{Tanaka} {et~al.}(2008){Tanaka}, {Mazzali}, {Benetti}, {Nomoto},
  {Elias-Rosa}, {Kotak}, {Pignata}, {Stanishev}, \& {Hachinger}}]{Tanaka08}
{Tanaka}, M., {Mazzali}, P.~A., {Benetti}, S., {et~al.} 2008, \apj, 677, 448

\bibitem[{{Townsend}(1958)}]{Townsend58}
{Townsend}, A.~A. 1958, Journal of Fluid Mechanics, 4, 361

\bibitem[{{Truran}(1982)}]{Truran82}
{Truran}, J.~W. 1982, in Essays in Nuclear Astrophysics, ed. C.~A. {Barnes},
  D.~D. {Clayton}, \& D.~N. {Schramm}, 467

\bibitem[{{Waldman} {et~al.}(2011){Waldman}, {Sauer}, {Livne}, {Perets},
  {Glasner}, {Mazzali}, {Truran}, \& {Gal-Yam}}]{Waldman11}
{Waldman}, R., {Sauer}, D., {Livne}, E., {et~al.} 2011, \apj, 738, 21

\bibitem[{{Webbink}(1984)}]{Webbink84}
{Webbink}, R.~F. 1984, \apj, 277, 355

\bibitem[{{Whelan} \& {Iben}(1973)}]{Whelan73}
{Whelan}, J., \& {Iben}, Jr., I. 1973, \apj, 186, 1007

\bibitem[{{Woosley} \& {Kasen}(2011)}]{Woosley11}
{Woosley}, S.~E., \& {Kasen}, D. 2011, \apj, 734, 38

\bibitem[{{Woosley} \& {Weaver}(1994)}]{Woosley94}
{Woosley}, S.~E., \& {Weaver}, T.~A. 1994, \apj, 423, 371

\bibitem[{{Yoon} \& {Langer}(2004{\natexlab{a}})}]{Yoon04b}
{Yoon}, S.-C., \& {Langer}, N. 2004{\natexlab{a}}, \aap, 419, 645

\bibitem[{{Yoon} \& {Langer}(2004{\natexlab{b}})}]{Yoon04}
---. 2004{\natexlab{b}}, \aap, 419, 623

\bibitem[{{Yoon} {et~al.}(2004){Yoon}, {Langer}, \& {Scheithauer}}]{Yoon04c}
{Yoon}, S.-C., {Langer}, N., \& {Scheithauer}, S. 2004, \aap, 425, 217

\bibitem[{{Zahn}(1992)}]{Zahn92}
{Zahn}, J.-P. 1992, \aap, 265, 115

\end{thebibliography}
\end{document}